\documentstyle[a4,12pt]{article} 
\input{epsf}            
\makeatletter

\font\tenmsy=msbm10
\font\sevenmsy=msbm10
\font\fivemsy=msbm10

\newfam\msyfam
\textfont\msyfam=\tenmsy  \scriptfont\msyfam=\sevenmsy
  \scriptscriptfont\msyfam=\fivemsy

\def\Bbb{\ifmmode\let\next\Bbb@\else
 \def\next{\errmessage{Use \string\Bbb\space only in math mode}}\fi\next}
\def\Bbb@#1{{\Bbb@@{#1}}}
\def\Bbb@@#1{\fam\msyfam#1}

\makeatother

\newcommand{\ZZ}{{\Bbb{Z}}}

\def\be{\begin{equation}}
\def\ee{\end{equation}}
\def\bea{\begin{eqnarray}}
\def\eea{\end{eqnarray}}

\begin{document}

\begin{titlepage}
{\hfill SWAT/102}
\vskip4cm
\centerline{\Large{\bf  U(1) Lattice Gauge theory and its Dual}}
\smallskip
\medskip
\centerline{\normalsize \bf P.~K.~Coyle$^{\rm a}$, I.~G.~Halliday$^{\rm b}$ and P. Suranyi$^{\rm c}$
}
\smallskip
\medskip

\begin{center}
{ \sl $^a$Racah Institute of Physics, 
 Hebrew University of Jerusalem, \\
Jerusalem 91904, Israel. \\
\medskip 
\sl $^b$Department of Physics, 
 University of Wales, Swansea \\
Singleton Park,
 Swansea, SA2 8PP, U.K. \\
\medskip 
 \sl $^c$Department of Physics, University of Cincinnati\\
Cincinnati, Ohio, 45221 U.S.A.\\
}\smallskip
\end{center}
\smallskip
\medskip
\abstract
The three dimensional  U(1) Lattice Gauge, in the weak coupling limit,
is dual to a Discrete Gaussian model. We  investigate this dual model 
and use it to calculate properties of the U(1) theory.
We find that, because of the nature of the dual model,
its  advantages  are outweighed by large
autocorrelation times generated when the dual system becomes disordered.

\end{titlepage}

\section{Introduction}

Duality transformations have provided a useful tool for investigating many
theories both in the continuum and on the lattice. The term duality has been used 
to describe many types of transformation. 
A common  property of these transformations is a mapping from the 
strong coupling region of one theory into the weak
coupling region
of its dual. Duality for lattice models was first developed by Kramers and
Wannier for the two dimensional Ising model \cite{kramers} and has since
been extended to other more complex lattice models \cite{kbm,savitRev}. 
Here we will use the
term duality to refer to the extension of this Kramers and Wannier duality. 

This type of transformation 
is appealing from a practical point of view.
Each model has its own set of advantages and disadvantages
making simulations difficult for some values of the coupling
while they become relatively easy to simulate in other regions.

By relating observables in the original theory to observables
in its dual it becomes possible to use the dual model 
for simulations where the original theory encounters
computational difficulties.  This is particularly relevant 
in the  weak coupling limit. Here the fields on the 
original lattice becomes ordered while the dual variables become 
increasingly disordered.

In this paper we investigate a compact U(1) lattice gauge theory
in three dimensions by performing simulations 
on its dual, the discrete Gaussian model. 
The duality transformation for 
this model
 was first performed
by Kogut et.al. \cite{kbm} and gives a model similar to the Solid on Solid 
models of
surface transitions, but in three dimensions.

\section{U(1) and the Discrete Gaussian Model}

We consider  a compact $U(1)$ gauge theory in 3 dimensions with
dynamical variables $U_{i,\mu}$ defined on links, where $i$ labels 
a site on the lattice and $\mu$ specifies the direction.  
Each link $U_{i,\mu}$ can
be expressed in terms of an angle 
$-\pi < \theta_{i,\mu} \le \pi$. 
\begin{equation} 
U_{i,\mu} = e^{i\theta_{i,\mu}}  
\end{equation}
with the Wilson action  defined as:
\begin{equation}
\beta S =  \beta  \sum_{\Box} (1- \cos(\theta_\Box) ) \label{eq:acti}
\end{equation}
where 
\bea
\theta_{i,\mu\nu} &=& \theta_{i,\mu}+\theta_{i+\hat{\mu},\nu}-\theta_{i+\hat{\nu},\mu}-\theta_{i,\nu}
\nonumber \\
         &=& \epsilon_{\mu\nu\gamma}\epsilon_{\gamma\alpha\beta}
\Delta_\alpha\theta_{i,\beta} \label{eq:theta}
\eea
and
\be
\theta_\Box = \frac{1}{2}\epsilon_{i,\gamma\mu\nu}\theta_{i,\mu\nu} 
\ee
The sum in equation (\ref{eq:acti}) is  over all sites $i$ and directions $\gamma$ 
to generate all the plaquettes on the lattice.
This gives a partition function:
\be
Z=\int_{-\pi}^{\pi}\prod_{j,\mu} d\theta_{j,\mu}  e^{-\beta \sum_\Box ( 1- \cos(\theta_\Box))}
\ee

Although this theory does
not experience a phase transition, and thus has no continuum limit,
it does possess an interesting topological structure. 
Through two duality like transformation, this partition function 
is  transformed
into the partition function for a coulomb gas of unbound
magnetic monopoles plus a free photon field \cite{poly,kbm,savitRev,savitU1,savit1}. 
This identifies the instantons as magnetic monopoles
which are exposed by writing
the plaquette angle (\ref{eq:theta}) as 
\be
\theta_\Box=\Theta_\Box+2\pi n_\Box \label{eq:theta-dec}
\ee
where $\Theta_\Box$ is in the range $-\pi<\Theta_\Box<\pi$ and 
$n_\Box = 0,\pm1,\pm2$.
This splits $\theta_\Box$ into a continuous part
and a part which cannot be written as a gradient.
One can therefore define a dual current
\be
j_{z-\hat{\mu},\mu}=\frac{1}{2}\epsilon_{\mu\nu\rho} n_{z,\nu\rho}
\ee
where $z$ is a site on the dual lattice. 
The monopoles are then identified by writing
\bea
\Delta_\mu j_{z,\mu} &=& m_z\\
m_z&=& 0,\pm1,\pm2,\pm3
\eea
When $m_z\ne 0$ there is a monopole/anti-monopole at the (dual) site
and the current $j_{z,\mu}$ exposes the Dirac strings connecting monopole
anti-monopole pairs. 
Dirac strings can also form closed loops without any monopoles.
With periodic boundary conditions these Dirac strings can form closed loops
which wrap around the boundary. The monopoles are physical objects which 
can be associated with cubes on the lattice however the Dirac strings
are gauge dependent so their
path can be altered through gauge transformations. 
Gauge transformations, however, cannot unwind loops which span the
boundaries so these loops
contribute to a  winding number which is gauge invariant.  
The  theory has only a confining `plasma'
phase for all values of $\beta$
although the monopoles are found to disappear from the system
above $\beta\approx 2.3$ \cite{3dtopo,landau}. 

\subsection{The Duality Transformation}
The dual model, exposed by Kogut et.al. \cite{kbm},
can be described by the partition function
\be
\label{eq:constraint}
Z= e^{-N\beta} 
\sum_{\{k\}}\exp\left(\sum_\Box\ln[I_k(\beta)]\right)
\prod_\Box\delta(\epsilon_{\mu\beta\alpha}\epsilon_{\alpha\gamma\rho}\Delta_\beta k_{i,\gamma\rho}) 
\ee
Where $k_{i,\gamma\rho}$ is the dual variable, $I_k(\beta)$ is the modified Bessel function,
N is the number of links on the lattice and
we have dropped the overall factors of $2\pi$ from the
delta functions.
As we explicitly sum over all values of $\theta_{i,\mu}$ to generate
equation (\ref{eq:constraint}) it implicitly includes all possible winding 
values.

By associating the $k$'s with links on the dual
lattice, rather than plaquettes of the original lattice,
equation
(\ref{eq:constraint}) can be interpreted as a spin theory
with the interactions on dual links given by 
\be
k_{z,\alpha} = \frac{1}{2}\epsilon_{\alpha\gamma\rho}k_{i,\gamma\rho}
\ee
where $z$ is the site dual to $i$.
The constraints on these dual interactions  then become
\be 
 \epsilon_{\mu\beta\alpha}\Delta_\beta k_{z,\alpha}=0 \label{eq:delta-dual}
\ee

\subsection{Solving the constraints \& Boundary Conditions}

As equation (\ref{eq:delta-dual}) is just the curl of $k$,
these constraints are automatically satisfied by writing
$k$ as the gradient of a scalar field.
Thus we can express the $k_{z,\alpha}$'s in terms of  a 
(dual) site variable $\phi_z$ 
\be
\label{eq:k_empty}
k_{z,\alpha}=\Delta_\alpha \phi_z 
\ee
where $\phi \in \ZZ$.
However we must be careful to include all the allowed $k_{z,\alpha}$
configurations. 
While equation (\ref{eq:k_empty}) guarantees the constraints are
satisfied it does not take into account effects of the periodic boundary
conditions.

For an infinite lattice we can write the $k_{z,\alpha}$  as a
potential  because of Stokes's Theorem. 
 This works for any closed loops within the
lattice.
 However if we have
periodic boundary conditions a contour crossing the boundary an odd
number of times forms a loop with no surface associated
with it. Stokes's Theorem therefore doesn't apply.
 Writing $k_{z,\alpha}$ as the gradient of $\phi$ forces this loop integral to be
zero however the constraints (\ref{eq:delta-dual}) do not impose this
restriction on $k_{z,\alpha}$. Therefore while equation (\ref{eq:k_empty}) is sufficient 
within the lattice, at the boundary we need a more general
prescription for $k_{z,\alpha}$ in order to generate all the allowed
configurations. 

For the boundary links we can write $k_{z,\alpha}$ as:
\be
\label{eq:k_boundary}
k_{z,\alpha}=\Delta_\alpha \phi_z + Q_{z,\alpha}(z) 
\ee
Using the constraints (\ref{eq:delta-dual}) on $Q$  gives
\be
\epsilon_{\mu\nu\sigma}\Delta_\nu(\Delta_\sigma \phi_z + Q_{z,\sigma}(z)) =0
\ee
Thus
\be
\Delta_\nu Q_{z,\sigma}(z) =0
\ee
for $\sigma \ne \nu$, so $Q_\sigma$ cannot change across each plane 
perpendicular to $\sigma$ leaving  only three extra degrees of
freedom $Q_i,Q_j$ and $Q_k$. 
Now if we consider the
line integral for a loop $C$ which crosses the $\mu$ boundary once we get
\bea
\int_{C} k_\mu \cdot dl &=& \sum_C k_\mu \\
&=& Q_\mu
\eea
These extra parameters therefore allow the fields $\phi$ to differ 
by  $Q$ across the boundary without any cost to the action.
This winding is the three dimensional extension of the
`step free energy' often introduced for two dimensional 
SOS models \cite{domb}. In two dimensions the cyclic boundary
conditions are related to pinning the boundary of the 
surface. The extra `step free energy', Q is introduced to remove this
pinning constraint in the hope of improved statistics.  

The minimum requirement for generating all the interactions
satisfying the constraints in (\ref{eq:constraint}) with 
periodic  boundary conditions is therefore
\be
k_{z,\mu} = \Delta_\mu \phi_z +Q_\mu
\ee
for links on the boundary and
\be
k_{z,\mu} = \Delta_\mu \phi_z 
\ee
for all other points.

As the step free energies $Q$ are 
related to the  boundary conditions they  become less important
as the lattice size is increased. 
When $\beta$ is small the fields
are predominantly at one level and, using the language of SOS 
models, the system is `smooth'.
Non-zero $Q$ values therefore increase the energy for every link 
on the boundary and are subsequently suppressed by a factor proportional 
to the lattice size squared. 
At large $\beta$ the system becomes `rough' with a `thickness'
that increases with $\beta$ and  with the lattice size. In this region 
changing $Q$
will, on average, decrease the energy for as many interactions as
it increases it, provided $Q$ is less than the thickness.
Correlation functions are unaffected by these changes in $Q$
and  with only three winding parameters, compared with the $L^3$ site
variables, fixing $Q$ should not influence correlation function
measurements, even for modest lattice sizes.
We therefore set $Q=0$ below although it can be 
replaced at any time if desired.
\be
\label{dual_Z}
Z=e^{-N\beta}\sum^{\infty}_{\{\phi\}=-\infty} \exp(\sum_{l_d}
\ln[I_{(\Delta_\rho \phi_z)}(\beta)])
\ee
Where we have a sum over configurations of site variables $\phi_z$
with an interaction defined on links $k_\rho=\Delta_\rho \phi_z$.

\subsection{The Weak Coupling limit}
By considering the weak coupling limit Kogut et.al were able to simplify
(\ref{dual_Z}) considerably. In this limit the partition function reduces
to a discrete Gaussian model. 
\be
\label{eq:approx}
\ln(I_n(\beta)) \approx -\frac{n^2}{2\beta} +\beta -\ln(2\pi\beta^{\frac{1}{2}})
\ee
To improve this approximation we can systematically add higher order
powers of $n$.
This approximation is equivalent to using the
periodic Gaussian or Villain action
\cite{villain}. The $\beta$ on the right had side of equation (\ref{eq:approx})
can therefore be identified with $\beta_v$, the Villain beta.
In his original paper Villain determined an approximate
relationship between the cosine beta, $\beta$ and $\beta_v$.
\be
\beta_v(\beta)=\left[2 \ln\left(\frac{I_0(\beta)}{I_1(\beta)}
\right)\right]^{-1} \label{eq:vil}
\ee
Where $I_0(\beta)$ and $I_1(\beta)$ are modified Bessel functions.
This relationship is found through matching the first  terms 
in the expansion of the cosine and Villain Boltzmann factors
and is therefore equivalent to adding extra terms 
in equation (\ref{eq:approx}).
A good approximation to  the partition function  is therefore 
\be
Z=\sum_{\{\phi\}} \exp\left(\sum_{\rho,j} \ln[\frac{I_0(\beta)}{I_1(\beta)}]
(\Delta_\rho \phi_j)^2\right)
\ee

\subsection{Wilson loops}

In the weak coupling limit Wilson loops become correlation functions
of the dual variables. 
Defining a loop
W(J) where the current $J_{i,\mu}$ defines the contour with  $J_{i,\mu}$
 given by:
\be
J_{i,\mu}= \left\{
\begin{array}{l}
+1 \mbox{ if link  } i \rightarrow i+\mu \mbox{ is in $C$. } \\
-1 \mbox{ if link } i+\mu \rightarrow  i \mbox{ is in $C$. } \\
0 \mbox{ otherwise }
\end{array} \right. \label{eq:wilsonJ}
\ee
The expectation value of this loop is then 
\be
W(J)=\frac{1}{Z}\prod\int_{-\pi}^{\pi}e^
{-\beta_v (\sum_{i,\Box}1-\cos(\theta_{i,\Box})+
 i\sum_{i,\mu}\theta_{i,\mu}J_{i,\mu})}
\ee
After the duality transformation, in the 
weak coupling limit, this become
\be
W(J)=\frac{1}{Z}\sum_{\{\phi\}}
e^{\frac{1}{2\beta_v}k^2_{\mu\nu}(\phi,S(J))} \label{eq:W-loop}
\ee
where in order to satisfy the new constraints 
$ k_{\mu\nu} $ is written as
\be
k_{\mu\nu}=S_{\mu\nu}
         +\epsilon_{\mu\nu\lambda}\Delta_\lambda \phi_i 
\ee
Where $S_\Box= 1$ for an oriented 
surface with $J_\rho\ne 0$ as its boundary and zero everywhere else. 
Expanding the exponential for the extra terms due to $S_\Box =1$
leaves
\be
W(J)=\frac{1}{Z}\sum_{\{\phi\}}
e^{\frac{A}{2\beta_v }} 
\left(
1 +\frac{(\sum_{l_d} l_{i,\rho} \Delta_\rho\phi_i)^2}{2\beta_v^2}
+\cdots \right)
e^{\frac{1}{2\beta_v}\sum_{l_d}(\Delta\phi_{i,\rho})^2}
\ee
where $A$ is the number of plaquettes in the surface 
$S_{\mu\nu}$ and the sum in
each term is over the dual links $l_\rho=\frac{1}{2}\epsilon_{\rho\mu\nu}S_{\mu\nu}$
which penetrate the surface $S_{\mu\nu}$.
Thus for the $1\times1$ Wilson loop, to order $\frac{1}{\beta_v^2}$ 
\be
 W(\Box) =
          e^{-\frac{1}{2\beta_v}}
         \left(1+{\frac{1}{\beta_v^2}}(
\langle\phi_i^2 -{\bar{\phi_i}}^2\rangle -
\langle\phi\phi_{i+1} -{\bar{\phi}}^2 \rangle) \right)
\label{eq:Wloop}
\ee
where $\bar{\phi}=\langle\phi\rangle$.

\section{Measurements}

In order to judge how useful the discrete Gaussian model is 
from a numerical perspective we also performed simulations using
a U(1) lattice measuring the $1\times1$ Wilson loop directly.
We considered lattices of size $8^3$ and $16^3$ with periodic 
boundary conditions. We used a standard metropolis algorithm 
and discarded the first 2,000 updates, where each update 
constituted a Metropolis hit for all links in the lattice.
50,000 updates were then performed taking measurements after
every update. The simulations were performed on DEC Alpha 3400AXP
workstations where,  for a $16^3$ lattice, each update took $0.212\pm0.001$
seconds.

Before attempting  accurate calculations for $U(1)$ observables using
the discrete Gaussian correlation functions  we first checked that
these measurements were insensitive to the step free energies.

We compared  simulations where the step free energy $Q_\alpha$ was
fixed at zero, with simulations where $Q_\alpha$ was dynamically updated
through a metropolis procedure  before each measurements. 
We found that $\langle |Q|\rangle$ remained zero for $\beta$ below $\approx 2.2$
while above it increased monotonically.
Measurements of the correlation functions however, were the same,
 within errorbars, for both static and dynamically varying 
step free energies. We therefore set $Q_\alpha=0$ for the remainder of
our simulations.

\begin{figure}[htbp]
  \begin{center}
    \leavevmode
\centerline{\hbox{\epsfxsize=14cm \epsfbox{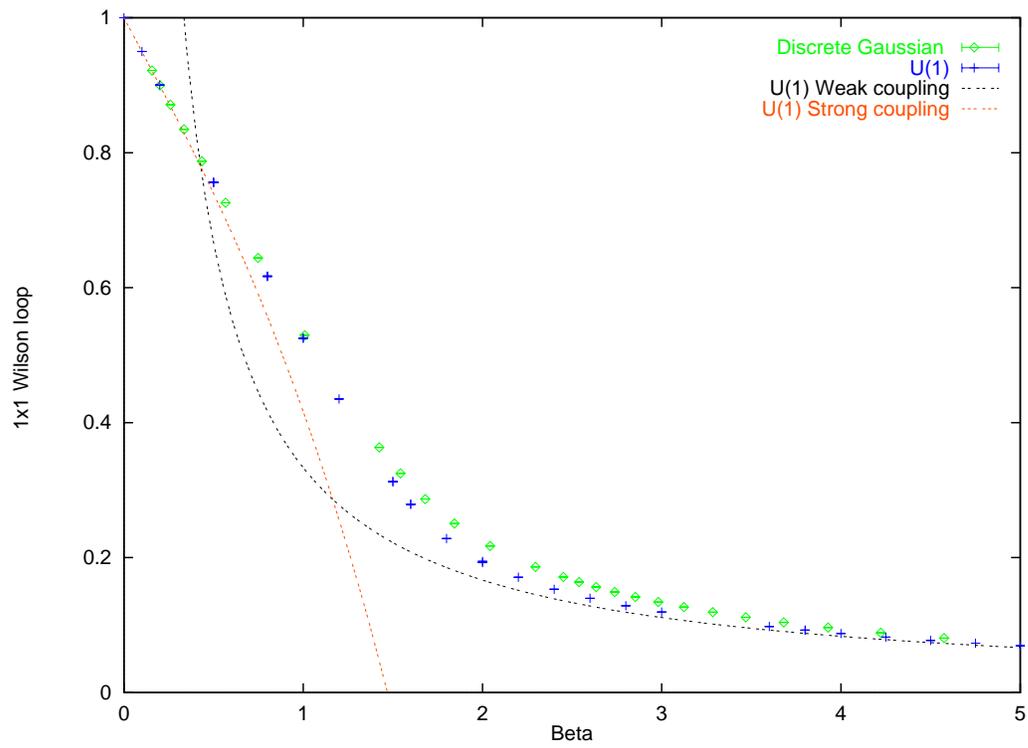}}}        
    \caption{$\langle\Box\rangle$ from U(1) and Discrete Gaussian simulations}
    \label{fig:sos-u1}
  \end{center}
\end{figure}

Simulations for the Discrete Gaussian model were  
performed on $8^3$ and $16^3$ lattices. Again, in order to
compare like with like, we implemented a metropolis algorithm
with each update consisting of a metropolis hit on each 
site of the lattice. We ran 210,000 updates and discarded the
first 10,000. Again we used DEC Alpha 3400AXP workstations
and each update took $0.033\pm0.001$ seconds for a $16^3$
lattice, almost 6.5 times faster than U(1).
We measured the correlation functions after
each update and used these to estimate the U(1) $1\times 1$
Wilson loop through equation \ref{eq:Wloop}. The results are
shown in figure \ref{fig:sos-u1} together with
 the  strong and weak coupling expansions
for the U(1) model  \cite{rothe,creutz-qgl}. 
\bea
\langle\Box\rangle &\approx& \frac{1}{3\beta} +\cdots
 \qquad \qquad \qquad\mbox{ to order } O(\frac{1}{\beta^2}) \\
                   &\approx& \frac{\beta}{2} + \frac{\beta^3}{12} +\cdots
 \qquad \mbox{ to order } O(\beta^3) 
\eea

\section{Errors and Efficiency}

In order to estimate the errors we split the raw data into bins of size 2 
to 500. We used this blocking to measure the overall errors
and calculate  auto-correlation times $\tau$. We also
checked that $\tau$ calculated in this way
was the same as $\tau_{exp}$ calculated from an 
exponential fit to the auto-correlation functions. 

\begin{figure}[htbp]
  \begin{center}
    \leavevmode
\centerline{\hbox{\epsfxsize=14cm \epsfbox{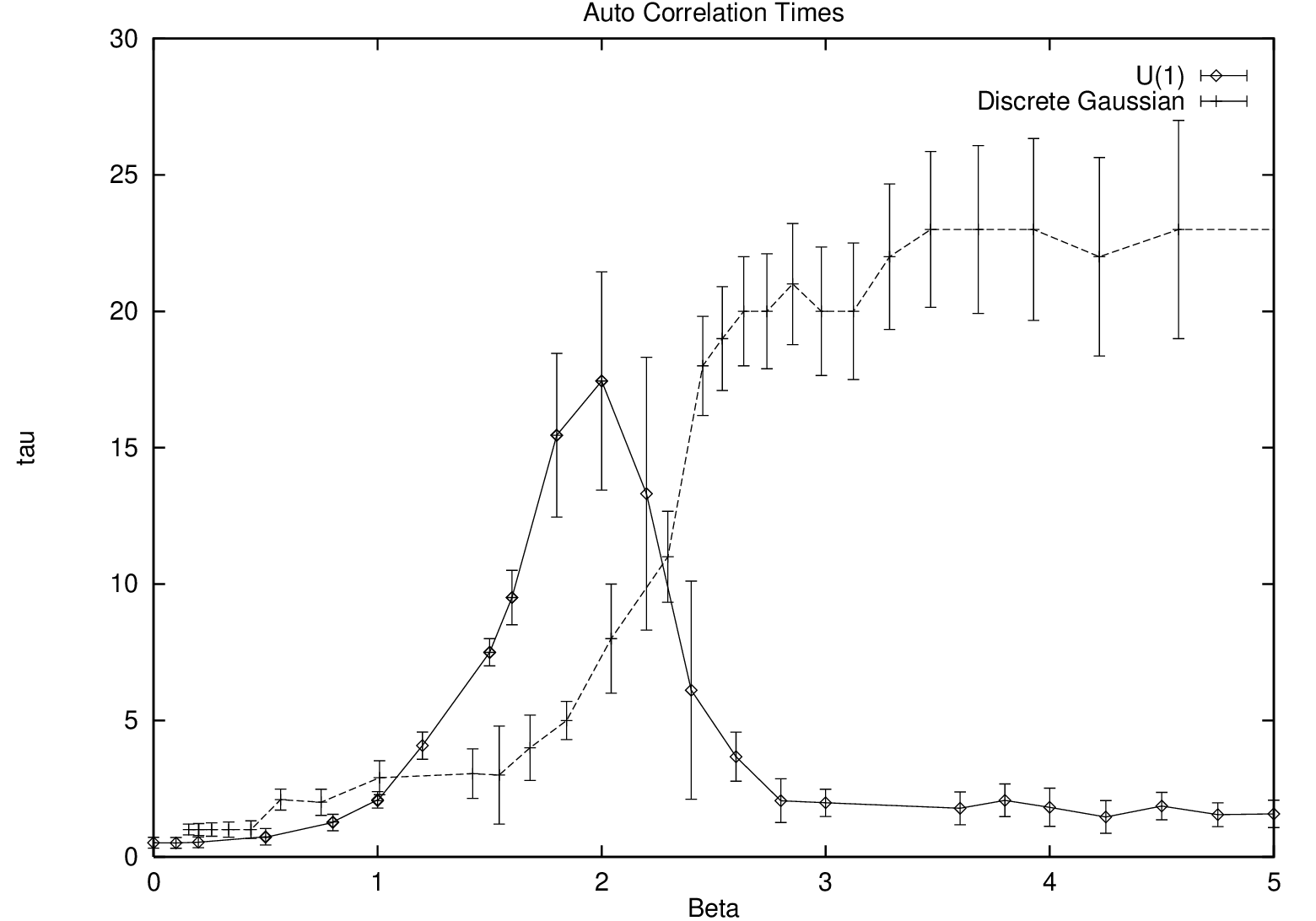}}}    
\vskip 1cm
\centerline{\hbox{\epsfxsize=14cm \epsfbox{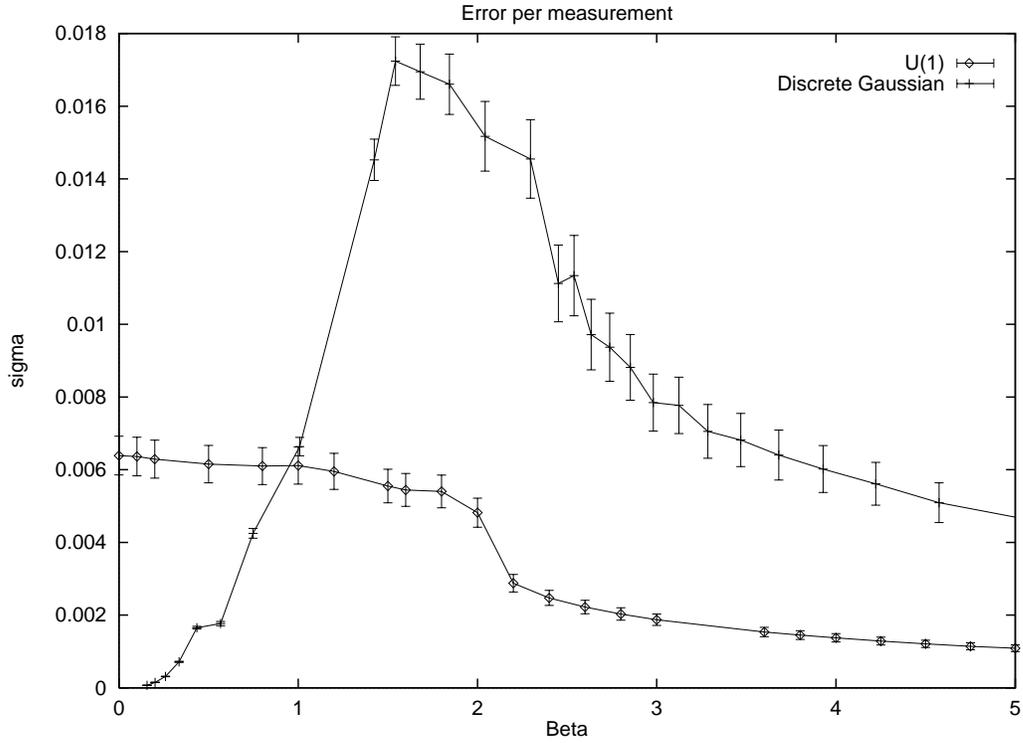}}}    
 \caption{ a) U1 Autocorrelations  Errors (algorithm error) 
           b) U1 Standard Errors }
    \label{fig:U1-s-error}
  \end{center}
\end{figure}

We used $\tau$ to split the errors in 
the calculation into two parts, an autocorrelation ($\tau$) 
component and a standard error ($\sigma$) component.
The total sample error,$\sigma_s$ for a given simulation is then
approximately
\be
\label{eq:errorN}
\sigma_{s} = \sigma\sqrt{\frac{2\tau}{ N}}
\ee
where N is the sample size.
The autocorrelation times measure the efficiency of the update
algorithm  and are  shown in figure \ref{fig:U1-s-error}a while
$\sigma$ measures the statistical errors for each 
independent measurement and depends only on the model.
These are shown in figure \ref{fig:U1-s-error}b where  
again we have used blocking to estimated errors.

For the U(1) lattice  the measurement error decreases
 significantly above $\beta \approx 2$. 
This is what we expect from the topological picture. At large
$\beta$ the monopoles  condense out and the effective 
number of degrees of freedom therefore reduces considerably. 
Autocorrelations also increase around $\beta\approx 2$ signaling the
emergence of extended objects. These are the monopoles
which become important as the temperature decreases before they
disappear from the system. Above $\beta\approx 2$ the autocorrelation
times fall when there are no monopoles. 
The autocorrelations remain short for large $\beta$.

For the Discrete Gaussian model autocorrelation times also 
increase around $\beta \approx 2$, however, they remain high
as $\beta$ increases. At small $\beta$ the lattice is smooth
but above  $\beta \approx 2$ it becomes rough and the local 
algorithm slows down.
This is  different to the usual slowing down in spin models
which occurs at low temperature. Here the large scale objects of the 
dual model appear at hight temperature as the dual model becomes more disordered
and the `thickness' increases. This causes the local metropolis
algorithm we used to have  difficulty updating configurations.
This suggest that a cluster algorithm 
may be effective at improving these simulations. A Solid On Solid
algorithm has been developed for the two dimensional discrete Gaussian
model by
Evertz et al. \cite{sorin-sos}. 
 This algorithm should be directly applicable to the three
dimensional model although, to our knowledge, it has not
been tested on a three dimensional lattice. 
The measurement error for the Discrete Gaussian model  experiences a peak around $\beta=2$ 
but most of this structure is generated by the transformation
in equation (\ref{eq:Wloop}).

\begin{figure}[htbp]
  \begin{center}
    \leavevmode
\centerline{\hbox{\epsfxsize=14cm \epsfbox{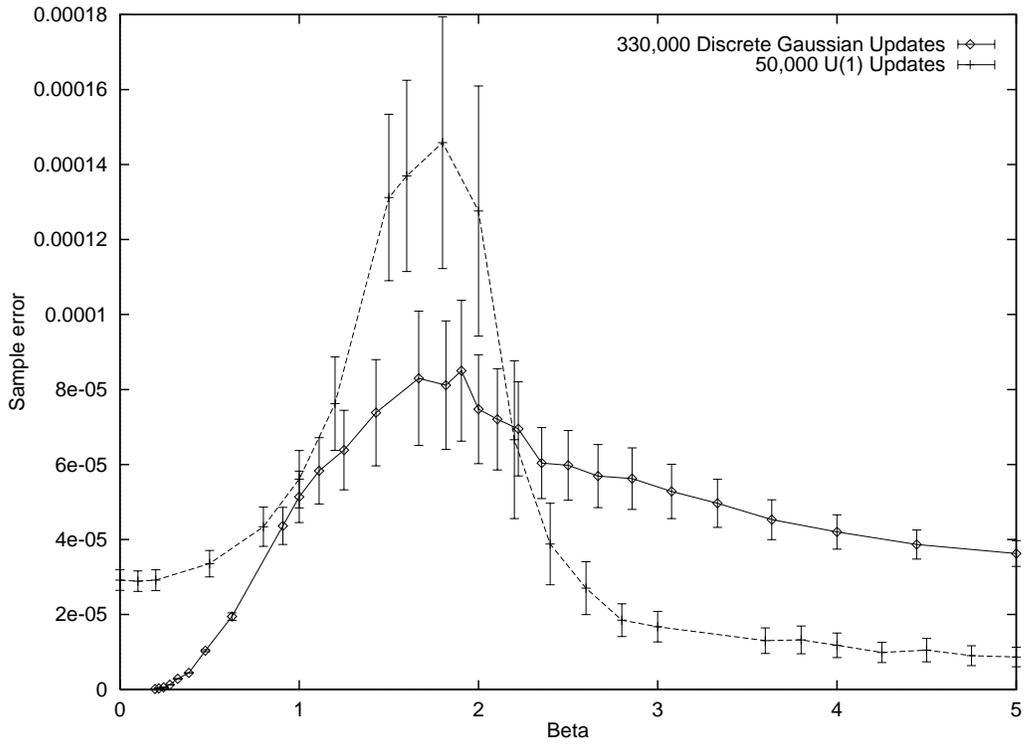}}}    
    \caption{Total error in $\langle S_{U1}\rangle$ for a given  
            CPU time. Errors are estimated using blocking}
    \label{fig:time}
  \end{center}
\end{figure}

Figure \ref{fig:U1-s-error} shows that both the auto-correlation times and the 
errors in each independent measurement are worse for the dual model
for most values of beta. However the dual simulations ran 
over six times faster than the equivalent U(1) calculations
so in order to compare the overall efficiency we need to consider
simulations requiring the same amount of CPU time.  
Using equation (\ref{eq:errorN})
we can  scale the results to estimate the errors for simulations 
taking the same amount of computer time.
This is shown in figure~\ref{fig:time}. The dual model  therefore only 
becomes more efficient for $\beta<2$. However as the transformation utilizes
the weak coupling limit the dual model is expected to be most
accurate for large $\beta$.

\section{Conclusion}

The main advantage of the dual model is the speed of its
updates compared to the equivalent U(1) lattice. This increase
in speed comes from using integer variables and a lack of local
gauge invariance.
This advantage however is destroyed by the large auto-correlation times
generated by the local metropolis algorithm. The Discrete Gaussian
 model has the unusual 
characteristic of generating large-scale structures when the 
fields become disordered. This is the region where the U(1)
system is at weak coupling and ordered and is when the transformation 
is most accurate.

Cluster algorithms exist for both the U(1) and  Discrete Gaussian 
 models. And the 
Discrete Gaussian simulations would have benefited dramatically from an efficient
algorithm. However our main concern was to study the 
transformation as a tool to improve the simulations. We were therefore
careful to consider both theories on an equal footing.
 Here we have seen 
that by studying the dual model the characteristics of the 
simulations can be changes significantly. 
However whether this is of any  practical use depends 
on the form of the dual model and the update algorithms available
in each case.

For U(1) it is unfortunate that the weak coupling region 
corresponds to the `rough' region of the dual model. 
This means that the disadvantages of both models 
occur when the duality transformation is most accurate. 
The opportunity to utilize better  statistical properties 
of the dual theory therefore does not occur and the only gains 
produced are from the smaller configuration space.
This is not always the case. Recently M.Zach et.al \cite{zach} 
have shown that in four dimensions, the U(1) dual model 
can be significantly more efficient than direct 
U(1) simulations.

The recent progress in duality transformations for 
non-Abelian models \cite{hal-sur1} allows for a similar
study of more complex dual theories. These more realistic models
transform to  dual theories without their 
large scale excitations at high dual temperatures.
They should therefore not encounter the simulation difficulties
experienced by the discrete Gaussian model. 
For U(1) in three dimensions, simulations are relatively efficient and leave 
little opportunity for improvement.  
The increased complexity and difficulties experienced with
 non-abilian models
allows more scope for improvement than this simple U(1) case.

\section*{Acknowledgments}
One of us (PS) would like to thank the University of Wales 
Research Opportunities Fund and the U.S. Department of 
Energy (Grant \#DE-FG02-84ER-40153) for financial support.
PKC acknowledges the financial support of a PPARC
research studentship.
\bibliography{thesis}
\end{document}